\documentclass{pasj01}
\Received{$\langle$reception date$\rangle$}
\Accepted{$\langle$acception date$\rangle$}
\Published{$\langle$publication date$\rangle$}
\usepackage{color}

\usepackage{wrapfig}

\begin{document}
\title{Formation of the Young Massive Cluster R136 triggered by Tidally-driven Colliding H{\sc i} Flows}

\author{Yasuo Fukui\altaffilmark{1, 2}, Kisetsu Tsuge\altaffilmark{2}, Hidetoshi Sano\altaffilmark{1, 2}, Kenji Bekki\altaffilmark{3}, Cameron Yozin\altaffilmark{3}, Kengo Tachihara\altaffilmark{2}, Tsuyoshi Inoue\altaffilmark{2}}%
\altaffiltext{1}{Institute for Advanced Research, Nagoya University, Furo-cho, Chikusa-ku, Nagoya 464-8601, Japan}
\altaffiltext{2}{Department of Physics, Nagoya University, Furo-cho, Chikusa-ku, Nagoya 464-8601, Japan}
\altaffiltext{3}{ICRAR, M468, The University of Western Australia, 35 Stirling Highway, Crawley Western Australia 6009, Australia}

\email{fukui@phys.nagoya-u.ac.jp, tsuge@phys.nagoya-u.ac.jp}

\maketitle

\begin{abstract}

Understanding of massive cluster formation is one of the important issues of astronomy.By analyzing the H{\sc i} data, we have identified that the two H{\sc i}  velocity components (L- and D-components) are colliding toward the H{\sc i} Ridge, in the southeastern end of the LMC, which hosts the young massive cluster R136 and $\sim$400 O/WR stars \citep{2013A&A...558A.134D} including the progenitor of SN1987A. The collision is possibly evidenced by bridge features connecting the two H{\sc i} components and  complementary distributions between them. We frame a hypothesis that the collision triggered the formation of R136 and the surrounding high-mass stars as well as the H{\sc i} \& Molecular Ridge. 
\citet{1990PASJ...42..505F} advocated that the last tidal interaction between the LMC and the SMC about 0.2 Gyr ago induced collision of the L- and D-components. This model is consistent with numerical simulations \citep{2007MNRAS.381L..16B}. We suggest that a dense H{\sc i} partly CO cloud of ~10$^{6}M_{\Sol}$, a precursor of R136, was formed at the shock-compressed interface between the colliding L- and D-components. 
We suggest that part of the low-metalicity gas from the SMC was mixed in the tidal interaction based on the $Planck/IRAS$ data of dust optical depth \citep{2014A&A...571A..11P}.
\end{abstract}

\section{Introduction} \label{sec:intro}

The formation mechanism of super star clusters (SSCs) is a subject of keen astronomical interest, because their extremely energetic interactions with surrounding material, in the form of UV radiation, stellar winds, and supernova explosions, are influential in driving galaxy evolution. In the Milky Way, we have thirteen SSCs as listed in a review article \citep{2010ARA&A..48..431P}, while the number of SSCs may increase by a more careful search for clusters in future. The young massive cluster RMC136 (R136), which is responsible for ionizing the largest H{\sc ii} region in the Local group (30~Dor), is estimated to have a mass of 10$^5 M_{\Sol}$, ten-times larger than SSCs in the Milky Way. Remarkably, R136 hosts the most massive stars over 200 $M_{\Sol}$ amongst all the SSCs known to date  \citep{2010MNRAS.408..731C}. The surroundings of R136 are also exceptionally active star forming regions involving $\sim$400 O/WR stars \citep{2013A&A...558A.134D}  associated with rich neutral hydrogen gas. The gas consists of the H{\sc i} Ridge, the Molecular Ridge and the CO Arc \citep{1992A&A...263...41L,1999PASJ...51..745F}, and its distribution is highly asymmetric in the LMC. 

It has been discussed that the two galaxies the LMC and SMC are interacting with each other in the past and the interaction is influencing the star formation history. \citet{1990PASJ...42..505F} presented a scenario in which the LMC and SMC had a close encounter ~0.2$\times$10$^9$ yrs ago, which perturbed the H{\sc i} gas in the both galaxies and caused the current highly asymmetric H{\sc i}/CO gas distribution in the LMC disk. These authors suggested that the perturbed gas collided with the LMC disk at 50--100 km s$^{-1}$ and triggered formation of R136 and the H{\sc i} Ridge with the highly asymmetric gas distribution. This scenario is supported by the numerical simulations of the galaxy interaction (e.g., \cite{2014MNRAS.443..522Y}), and is consistent with that the H{\sc i} consists of two velocity features, the L- and D-components (\cite{1992A&A...263...41L}; see a typical spectrum in Figure 1a), whereas a detailed observational test of the scenario was not undertaken to date. 

The aim of the present paper is to analyse the high-angular resolution The Australia Telescope Compact Array (ATCA) +Parkes H{\sc i} and The Magellanic Mopra Assesment (MAGMA) CO data in order to provide observational evidence for collision between the L- and D-components and to test a scenario of formation of R136 based on cloud-cloud collision (CCC) models (\cite{2009ApJ...696L.115F,2010ApJ...709..975O}; \cite{2011ApJ...738...46T}; \yearcite{2015ApJ...806....7T,2017ApJ...835..142T}; \cite{2014ApJ...780...36F}; \yearcite{2015ApJ...807L...4F,2016ApJ...820...26F,2017arXiv170104669F}; \cite{2017ApJ...835..108S}).


\section{Observational data and masking} \label{sec:obs}

The ATCA +Parkes H{\sc i} 21 cm data \citep{2005astro.ph..6224K} are used in this study. The resolution of the combined H{\sc i} image is 1$\farcm$0. This is 15 times higher resolution than that of \citet{1992A&A...263...41L}, which showed that the H{\sc i} distribution consists of the L- and D-components. The rms noise fluctuations of the combined map, determined from the line-free parts of the final data cube, is  15 mJy beam$^{-1}$. This corresponds to a brightness temperature sensitivity of  2.4 K (1$\sigma$), for velocity resolution of 1.649 km s$^{-1}$.

We used the data of the rotatinal transition of $^{12}$CO($J$ = 1--0) observed with NANTEN 4 m telescope (\cite{1999PASJ...51..745F}; \yearcite{2008ApJS..178...56F}; \cite{2009gcgg.book..121K}; for a review of molecular clouds in 
the LMC and SMC see \cite{2010ARA&A..48..547F}) for a large scale analysis. The half-power beamwidth is 2$\farcm$6, and the velocity resolution is 0.65 km s$^{-1}$. 
We also used the MAGMA $^{12}$CO($J$=1--0) data \citep{2011ApJS..197...16W} for a detailed analysis. The spatial resolution is $\timeform{45''}$, and the velocity resolution is 0.53 km s$^{-1}$. 

The archival data set of dust optical depth at 353 GHz, $\tau _{353}$, and dust temperature, $T_{d}$ obtained by the $Planck$ and $IRAS$ data (see \cite{2014A&A...571A..11P} for detail) are used to make comparisons with the H{\sc i} data.  
We also used  the data from the Southern H--Alpha Sky Survey Atlas (SHASSA) at 656.3 nm wavelength, the H$\alpha$ emission line of hydrogen \citep{2001PASP..113.1326G}. The angular resolution is about 0$\farcm$8, and the sensitivity level is 2 Rayleigh (1.2$\times 10^{-17}$ erg cm$^{-2}$ s$^{-1}$ arcsec$^{-2}$) pixel$^{-1}$.

We masked the data in order to eliminate regions with locally heated components and those with different environment as indicated by their high density, by using the H$\alpha$ and CO data when we compare $\tau_{353}$ and H{\sc i}; the areas showing CO emission higher than 1 K km s$^{-1}$ (1$\sigma$) and the areas showing H$\alpha$ emission higher than 30 Rayleigh are masked \citep{2014ApJ...796...59F, 2015ApJ...798....6F,okamoto2017inprep}. 

\section{Results} \label{sec:results}
\begin{figure*}[htbp]
\begin{center}
\includegraphics[width=16cm]{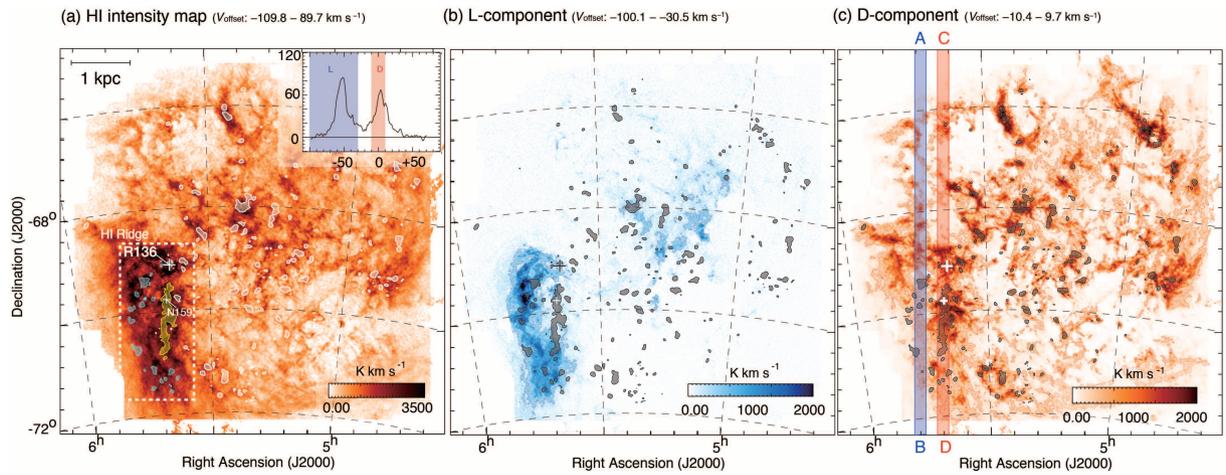}

\end{center}
\caption{(a) The H{\sc i} integrated intensity map with the integration velocity range of  $V_\mathrm{offset}$ = $-100.1$--$89.7$ km s$^{-1}$. The white dashed box illustrate the H{\sc i} Ridge. The upper right panel shows a typical spectrum of H{\sc i} at (R.A., Dec.) = ($\timeform{5h46m43.54s}$,$\timeform{-69D42'59.31''}$). The horizontal axis and vertical axis indicate $V_\mathrm{offset}$ [km s$^{-1}$] and intensity [K], respectively. A blue shaded range and a red shaded range indicate the integration velocity range of L- and D-components, respectively. (b) The H{\sc i} integrated intensity maps of L-component with a velocity range of $-100.1$--$-30.5$ km s$^{-1}$. The contour levels are 350, 800, 1200, and 2400 K km s$^{-1}$. (c) Same as (b) but for D-component with a velocity range of $-10.4$--$9.7$ km s$^{-1}$ and the contour levels of 400, 800, 1200, and 1600 K km s$^{-1}$. The crosses in (b) and (c) denote the positions of R136 and LHA 120-N 159 (N159), while the shaded areas in the 3 panels delineate the regions with CO intensity greater than 3 $\sigma$ with the same velocity range. Those in yellow and blue in (a) indicate the Molecular Ridge and the CO Arc, respectively. The blue and red lines in (c) show the integration ranges in R.A. in Figure 2.}
\label{fig1}
\end{figure*}%
Following the method of \citet{1992A&A...263...41L} we subtracted the galactic rotation 
in Figure 1, and reproduced their results with 1$\farcm$0 resolution. Figure 1 shows the new images of the two H{\sc i} components along with the CO clouds. The main feature of the L-component is located toward the CO Arc and in the east of the Molecular Ridge, while the D-component is extended over the galactic disk. We confirmed that the asymmetric distribution of H{\sc i} in Figure 1a is caused by adding the L-component.

Figure 2 shows two Dec.-velocity diagrams along the CO Arc and the Molecular Ridge. In Figure 2 we find the two velocity components and the bridge features between them. The intensity of the bridge features at each Dec. is correlated with the CO clouds and with the enhanced H{\sc i} intensity, suggesting collisional interaction in CCC indicated by numerical simulations (\cite{2013ApJ...774L..31I,2014ApJ...792...63T,2015MNRAS.450...10H}; \yearcite{2015MNRAS.454.1634H}).

\begin{figure}[htbp]
\begin{center}
\includegraphics[width=\linewidth]{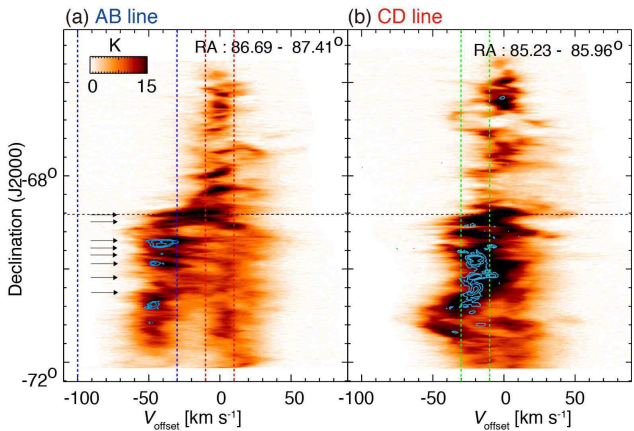}
\end{center}
\caption{Dec.-velocity diagrams of H{\sc i} superposed on the CO contour. The position of R136 is indicated by the dashed horizontal line. The contour levels are 0.015, 0.03, and 0.05 K km s$^{-1}$ for each panel.  The integration ranges in the R.A. are from 86$\fdg$69 to $87\fdg41$ in (a), and $85\fdg23$ to $85\fdg96$ in (b), respectively. The blue and red dashed vertical lines indicate the velocity range of L- and D-components, respectively. The black arrows in (b) indicate the position of bridge features. The green dashed vertical lines indicates intermediate velocity range of L- and D-components.}
\end{figure}%

Figure 3 shows an overlay of the L- and D-components in the region including 30 Dor. We see the two components show complementary distribution; in Figure 3a the D-component shows intensity depression toward the dense part of the L-component, and the L-component shows depression toward the dense part of the D-component. 
There are two places of the complementary distribution in the H{\sc i} Ridge; one is toward (R.A., Dec.)$\sim$($\timeform{5h40m0s}$--$\timeform{5h54m0s}$, $\timeform{-68D30'0''}$--$\timeform{-70D12'0''}$) and the other (R.A., Dec.)$\sim$($\timeform{5h38m0s}$--$\timeform{5h51m0s}$, $\timeform{-70D12'0''}$--$\timeform{-71D30'0''}$). 


We find the complementary distribution between the L- and D-components shows some displacement. This displacement between the two complementary features suggests that the relative motion makes a large angle to the line-of-sight \citep{2017arXiv170104669F}. We applied the the overlapping function $H$($\Delta$) in pc$^{2}$ to calculate the displacement in the complementary distribution which is presented in the Appendix of \citet{2017arXiv170104669F} 
in such a way that the two features coincide spatially after the displacement. Specifically, we derived the projected displacement where the overlapping area of the strong L-component (intensity larger than 1000 K km s$^{-1}$) and the depression of the D-component (intensity smaller than 550 K km s$^{-1}$) becomes maximum.
In Figure 3b we see two H{\sc i} clouds in the L-component, the north cloud at (R.A., Dec.)$\sim$($\timeform{5h38m0s}$--$\timeform{5h50m0s}$,$\timeform{-68D20'0''}$-- $\timeform{-69D48'0''}$) and the south cloud at ($\timeform{5h36m0s}$--$\timeform{5h49m0s}$,$\timeform{-70D6'0''}$-- $\timeform{-71D20'0''}$), which are surrounded by the D-component. They exhibit a complementary distribution with displacements of 170 pc and 260 pc lengths as shown by the arrows, respectively. We confirmed that the position angle 320 deg. gives the best fit by changing the angle from 315 to 325 deg. 

Figure 3c shows a closeup view in the R136 region, showing complementary distributions between the two velocities of H{\sc i} at $\sim$100pc scale; the two blue-shifted H{\sc i} intensity depressions at (R.A., Dec.) = ($\timeform{5h40m0s}$--$\timeform{5h42m0s}$,$\timeform{-68D50'0''}$-- $\timeform{-69D00'0''}$) and ($\timeform{5h38m0s}$--$\timeform{5h42m0s}$,$\timeform{-69D04'0''}$-- $\timeform{-69D20'0''}$) are clearly associated with the H{\sc i} components toward the same position, where the H{\sc i} distribution within 200 pc of R136 is not reliable due to artificial effects of the interferometeric fringes.

\begin{figure*}
\begin{center}
\includegraphics[width=16cm]{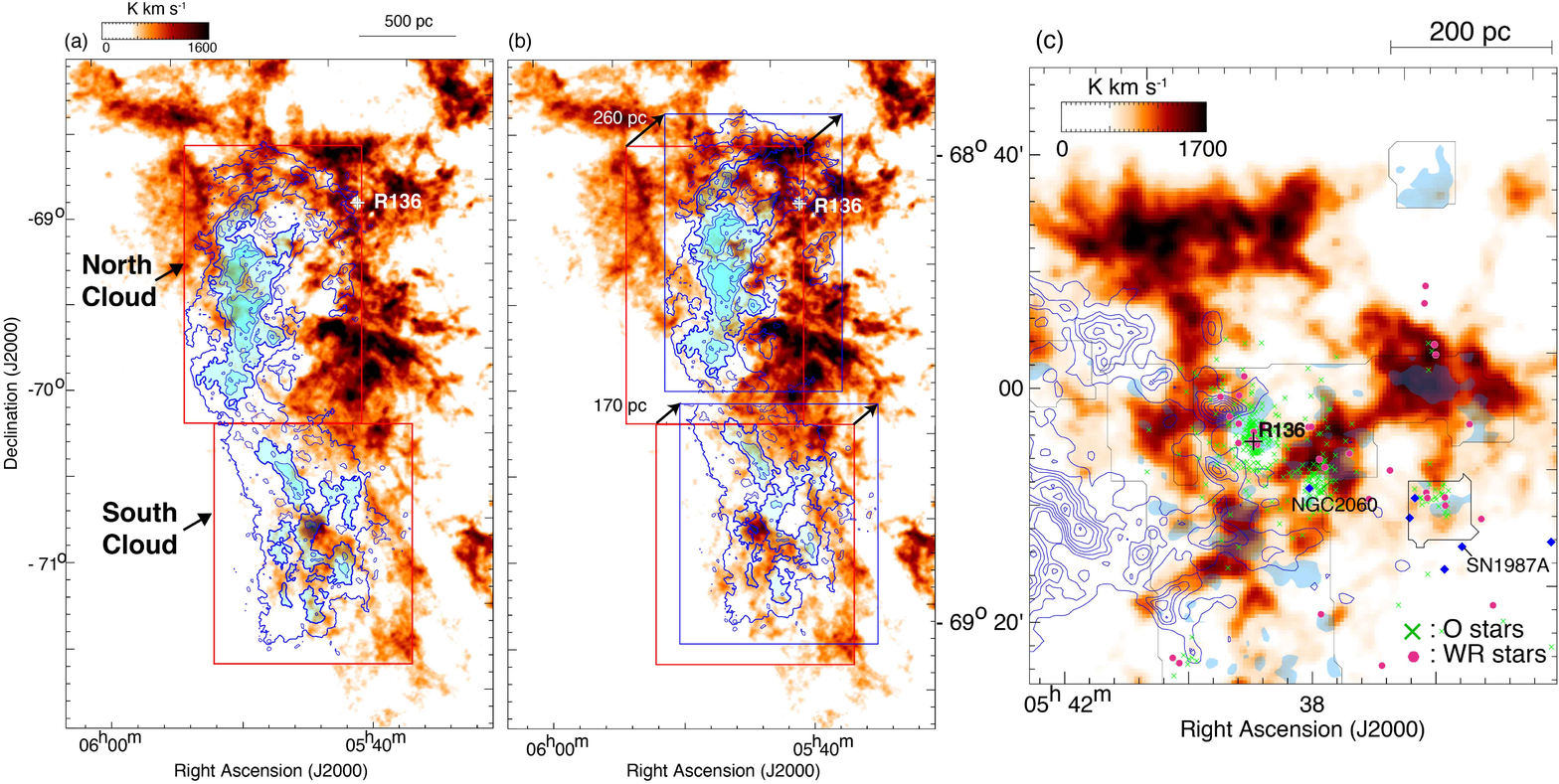}

\end{center}
\caption{(a) H{\sc i} intensity map of D-component superposed on the L-component's contours of the CO Arc and the Molecular Ridge . The contour levels are 500, 800, 1000, 1200, 1400, 1600, 1800, and 1900 K km s$^{-1}$.  (b) Same image as (a), whereas contours of L-component are displaced. The projected displacements of the northern half and the southern half are 260 pc and 170 pc, respectively, at a position angle of 320 deg. These distances are determined by the overlapping integral function $H$($\Delta$) (see the text and the Appendix of \cite{2017arXiv170104669F} ). The red boxes indicate the initial position of L-component, and the blue boxes of (b) indicate the displaced position. Blue shaded regions of (a) and (b) indicate the H{\sc i} intensity is greater than 800 K km s$^{-1}$. (c) H{\sc i} intensity map of D-component superposed on the contours of L-component toward R136. The integration velocity ranges of L- and D-components are the same as Figure \ref{fig1} and the contour levels are 400, 600, 700, 800, 900, 1100, 1150, 1200, and 1250 K km s$^{-1}$.  The L-component's contours are not displaced. Blue shaded regions indicate the CO intensity of 3$\sigma$ at the same velocity range with Figure 1(a). Black cross, green crosses, and pink circles indicate the position of R136, O stars, and WR stars respectively. Blue diamonds indicate the positions of SNRs \citep{2016A&A...585A.162M}.}
\label{fig}
\end{figure*}%

Figure 4 shows a scatter plot between $W$(H{\sc i}) and $\tau_{353}$, where the galactic foreground component is subtracted by using the $W$(H{\sc i})--$\tau_{353}$ relation \citep{2015ApJ...807L...4F}. The plots for the D-component and the H{\sc i} Ridge show significantly different slopes by a factor of 2. The difference in the slope corresponds to the difference in metalicity, if we assume that the dust-to-metal ratio is constant in the LMC.

\begin{figure}
\begin{center}
\includegraphics[width=\linewidth]{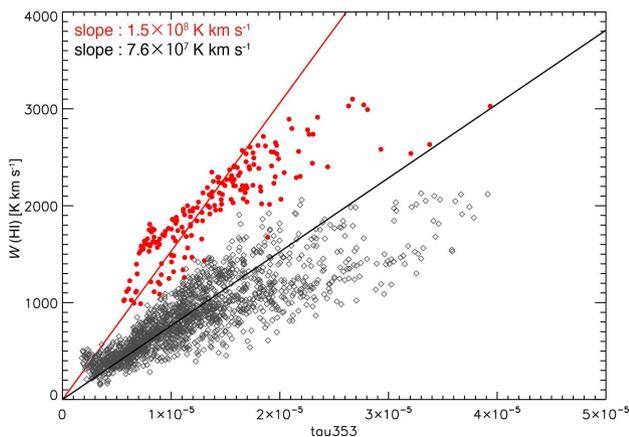}
\end{center}
\caption{Scatter plot between $\tau _{353}$ and $W$(H{\sc i}) for $T_{\rm d}$ $>$ 22.5 K.  Red points are the data of Arc and Molecular Ridge (the region of Figure 3a), and Grey points are the data of the optical stellar bar region. The red and black linear lines are obtained by the least-squares fit which is assumed to have zero intercept. The data of the direction of the optical stellar bar are used for fitting of the black line, and the red data are used for fitting of the red line.}
\end{figure}%

\section{Discussion} \label{sec:discussion}
\subsection{Cloud-cloud collision model} \label{sec:CCC}
The present results indicate that the L- and D-components show observational signatures of cloud-cloud collision, i.e., the two velocity components, the bridge features between them and the complementary distributions. \citet{2016ApJ...820...26F}  summarize and discuss these properties based on observations of ten cases of O star formation, eight in the Milky Way and two in the LMC, and hydrodynamical numerical simulations of cloud-cloud collisions. The differences between the ten previous cases and the present one in the LMC are that the velocity separation 50 km s$^{-1}$ and the size of the colliding components 2 kpc are significantly larger than in the previous cases, and that the O and WR stars including R136 are considerably more numerous than the previous cases ($\sim$400 in total over an area of 200 pc $\times$ 200 pc) \citep{2013A&A...558A.134D}. We see two outstanding stellar condensations in R136 and NGC~2060 (Figure 3c). 


The major overlapping of the two components in the southeast (Figure 1) is called the H{\sc i} Ridge. H{\sc i} intensity of H{\sc i} Ridge is elevated above a contour level 2000 km s$^{-1}$. By the optically thin approximation the $N$(H{\sc i}) in the ridge is estimated to be (0.4--$1)\times10^{22}$ cm$^{-2}$ on a 20 pc scale. The H{\sc i} Ridge includes two major elongated CO clouds, the Molecular Ridge and the CO Arc. The total mass of the H{\sc i} Ridge is estimated to be 1$\times 10^{8} M_{\solar}$ and the molecular masses of the Molecular Ridge and the CO Arc are 1$\times10^{7}M_{\Sol} $and 0.5$\times 10 ^{7}M_{\Sol}$, respectively \citep{2001PASJ...53..971M}.
A scenario in which the two velocity H{\sc i} components were created by the tidal interaction between the LMC and SMC 2$\times$10$^8 $ yrs ago was presented by \citet{1990PASJ...42..505F}. 
Subsequently, numerical simulations of the galactic interaction and the gas motion were carried out by \citet{2007PASA...24...21B} and (\yearcite{2007MNRAS.381L..16B}). 
The present analysis shows that the stripped H{\sc i} gas is at present falling onto the LMC disk at a projected velocity of 50 km s$^{-1}$, as is consisted with the numerical simulations by \citet{2007MNRAS.381L..16B}. This interpretation indicates that L-component was on the far side of the LMC disk and is now nearly at the same location with D-component, as suggested by the ongoing collisional interaction.

\subsection{Ongoing collision} \label{sec:ongoing}
A possible scenario for the collisional interaction is framed as follows, 
\begin{enumerate}
\item \underline{The region immediately surrounding R136} has a high $N$(H{\sc i}) of $\sim$10$^{22}$ cm$^{-2}$. The other regions of O/WR star formation or O/WR stars are mostly distributed only on the western side of R136, where the parent gas is already strongly dispersed by the stellar ionization.  They include the region of the super shell 30 Dor C, the progenitor of SN1987A, etc., which is exposed to us, where we see some CO and H{\sc i} gas on small scales. 
The CO gas in the R136 region is strongly dispersed as shown by the recent ALMA results \citep{2013ApJ...774...73I} whereas it is still associated with massive clumpy H{\sc i} gas (Figure 2). The age of R136 is estimated to be 1.5 Myrs \citep{2016MNRAS.458..624C} and the other O/WR stars are probably older than R136 with ages of more than a few Myrs as suggested by the relative paucity of ISM compared with the R136 region. 
The vicinity of 
R136 shows an ionized H{\sc i} cavity of a 15 pc radius (Figure 3c). 
The formation of the O/WR stars took place some Myrs ago and R136 was most recently formed. By considering the accumulated empirical knowledge \citep{2017arXiv170104669F}, we suggest that the distribution of these high mass stars reflects the initial density distribution of the H{\sc i} gas prior to the collision; very dense 
compact gas formed R136 by collision and less dense gas 
formed extended O/WR stars . The masses of the two H{\sc i} clumps having 100 pc size in Figure 3c are estimated to be ~$1\times10^{6}$ $M_{\solar}$. This is consistent with the mass of R136 $1\times10^{5}$ $M_{\solar}$ which requires a parent cloud of $1\times10^{6}$ $M_{\Sol}$ for an assumed star formation efficiency. 

\item \underline{The CO Arc region} shows two velocity components separated by 50 km s$^{-1}$, which show clear bridge features as well as complementary distributions typical to collision (Figure 2a). These features indicate that the collision initiated 2 Myrs ago as roughly estimated from a ratio of the cloud size and the velocity 100 pc/50 km s$^{-1}$. The L-component shows a little sign of deceleration, probably because $N$(H{\sc i}) of the L-component dominates that of the D-component toward the regions; $N$(H{\sc i}) of the L-component is $\sim$4$\times 10^{21}$ cm$^{-2}$ while that of the D-component is $\sim1\times10^{21}$ cm$^{-2}$. This difference results in little deceleration of the L-component by momentum conservation. The bridge features support the collisional interaction as shown by numerical simulations \citep{2014ApJ...792...63T,2015MNRAS.454.1634H}. It is possible that the low average column density in the D-component did not accommodate formation of massive dense clumps in which O~stars can form.

\item \underline{The Molecular Ridge region} shows a single velocity broad feature in Figure 2b with the CO clouds. Their peak velocity lies between L- and D-components after subtraction of the galactic rotation. We suggest that since $N$(H{\sc i}) of the two components was nearly the same, the collision resulted in merging of the two components at their intermediate velocity via momentum conservation. 
Then, total $N$(H{\sc i}) of the merged component was elevated to $\sim$1$\times10^{22}$ cm$^{-2}$. The Molecular Ridge was formed by the strong compression in the collision, while some of the CO gas was formed prior to the collision. 
In N159 E and W we found collision between filamentary clouds with a velocity of 7--8 km s$^{-1}$ by ALMA observations at $\timeform{1''} $resolution \citep{2015ApJ...798....6F,2017ApJ...835..108S}. It is speculated that the highly filamentary distribution in N159 reflects the gas dynamics in the merging interface layer where filaments are formed according to numerical simulations \citep{2013ApJ...774L..31I}.

\end{enumerate}

\subsection{GMC/cluster formation} \label{sec:GMC}
The collision of H{\sc i} gas flow is able to compress the H{\sc i} gas and significantly increases the density according to MHD numerical simulations \citep{2012ApJ...759...35I} . H$_{2}$ clouds are formed in the interface layer in $\sim$Myr for density of 10$^{3}$ cm$^{-3}$, and it is expected that the molecular clouds become self-gravitating if they become massive enough. The subsequent process is calculated by \citet{2013ApJ...774L..31I} which simulates the colliding flow of dense molecular gas by incorporating self-gravity. These simulations indicate formation of dense clumps which can directly lead to high-mass star formation 
on smaller scales of 10--20 pc. In a kpc scale collision in the LMC we infer that part of the dense H{\sc i} gas may be converted into Giant Molecular Clouds (GMCs) if H{\sc i} gas are decelerated by the collision and further compressed. \citet{2009ApJ...705..144F} presented a scenario of formation of GMCs in the LMC via gravitational H{\sc i} accretion by showing the H{\sc i} becomes denser toward GMCs based on the H{\sc i} data \citep{2003ApJS..148..473K}. 


Figure 4 shows a scatter plot between the dust optical depth at 353 GHz $\tau_{353}$ and the 21 cm H{\sc i} intensity $W$(H{\sc i}) in the LMC, where only the highest dust temperature components are shown. We assumed that the highest dust-temperature components ($T_{d}$ > 22.5 K) are most probably optically thin . The overall distribution of the plot is well represented by linear regressions in Figure 4. The points in the H{\sc i} ridge are shown by red color and show a slope by a factor of 2 steeper than the rest of the points outside the H{\sc i} Ridge. We interpret that this difference is due to different dust abundance between them; the H{\sc i} ridge has significantly less dust grains than the rest of the LMC. The numerical simulations by \citet{2007MNRAS.381L..16B} shows that the H{\sc i} gas of the SMC is possibly mixed with the LMC gas in the tidal interaction. If this happens, the H{\sc i} ridge contains a significant amount of low-metalicity H{\sc i} of the SMC. The metalicity of the SMC is smaller than that of the LMC by a factor of five \citep{2002A&A...396...53R}. If the H{\sc i} ridge consists of H{\sc i} mass both from the SMC and LMC at a ratio of ~1:1, for sub-solar metalicity 1/10 in the SMC and the 1/2 in the LMC, a factor of 2 different slopes in Figure 4 is explained, lending a support for the tidal origin of the H{\sc i} ridge (\cite{2014ApJ...780...36F}; \yearcite{2015ApJ...807L...4F}).

\section{Conclusions} \label{sec:conclusion}
We found signatures of a collision over a few kpc between the two velocity H{\sc i} components toward R136, the H{\sc i} Ridge including the Molecular Ridge \& the CO Arc in the LMC. 
The blue-shifted H{\sc i} L-component was created in the galaxy interaction 0.2 Gyrs ago via tidal stripping between the LMC and the SMC. In 0.2 Gyrs, the perturbed H{\sc i} gas 
is now colliding interacting with the H{\sc i} gas in the LMC disk. 
The collision formed $\sim$ 400 O/WR stars in the H{\sc i} Ridge which includes R136, N159 and the other active star forming GMCs, alongside the CO Arc, as supported by numerical simulations of colliding H{\sc i} flows. 


\begin{ack}
This work was financially supported in part by JSPS KAKENHI Grant Numbers 16K17664 and 16H05694. This work was also financially supported by Career Development Project for Researchers of Allied Universities. 
The ATCA, Parkes and Mopra radio telescope are part of the ATNF which is funded by the Australian Government for operation as a National Facility managed by CSIRO. The UNSW Digital Filter Bank used for the observations with the Mopra Telescope was provided with support from the Australian Research Council. Based on observations obtained with Planck, an ESA science mission with instruments and contributions directly funded by ESA Member States, NASA, and Canada. The Southern H-Alpha Sky Survey Atlas, which is supported by the National Science Foundation.
\end{ack}

\end{document}